\documentclass[pre,superscriptaddress,twocolumn,showpacs,showkeys,preprintnumbers,amsmath,amssymb,floatfix]{revtex4}

\usepackage{graphicx}
\usepackage{dcolumn}
\usepackage{bm}

\begin{document}

\title{Spectral methods cluster words of the same class in a syntactic dependency network}

\author{Ramon \surname{Ferrer i Cancho}}
\affiliation{INFM UdR Roma1, Dip. di Fisica, Universit\`a 
``La Sapienza''. Piazzale A. Moro 5, 00185 Roma, Italy.}
\author{Andrea Capocci}
\affiliation{Centro Studi e Ricerche e Museo della Fisica,
  ``E. Fermi'', Compendio Viminale, Roma, Italy.}
\author{Guido Caldarelli}
\affiliation{INFM UdR Roma1, Dip. di Fisica, Universit\`a 
``La Sapienza''. Piazzale A. Moro 5, 00185 Roma, Italy.}
\affiliation{Centro Studi e Ricerche e Museo della Fisica,
  ``E. Fermi'', Compendio Viminale, Roma, Italy.}

\date{\today}

\begin{abstract}
We analyze here a particular kind of linguistic network where vertices represent
words and edges stand for syntactic relationships between words. The statistical
properties of these networks have been recently studied and various features
such as the small-world phenomenon and a scale-free distribution of degrees have
been found.  Our work focuses on four classes of words: verbs, nouns, adverbs and
adjectives.  Here, we use spectral methods sorting vertices. We show that the
ordering clusters words of the same class.  For nouns and verbs, the cluster
size distribution clearly follows a power-law distribution that cannot be explained
by a null hypothesis.  Long-range correlations are found between vertices in the
ordering provided by the spectral method.  The findings support the use of
spectral methods for detecting community structure.
\end{abstract}



\maketitle 

\section{Introduction}

\label{introduction_section}
A great amount of efforts has been recently devoted to the study of complex
networks \cite{Barabasi2001a,Buchanan2002a}. The topology of systems as
different as the Internet \cite{Caldarelli2000a,PastorSatorras2004a}, the World
Wide Web \cite{Adamic1999a,Adamic2000a,Albert1999}, biological
\cite{Vazquez2003a} and social systems\cite{Watts1999} have been found to share
similar statistical properties.  Many networks display both a distribution of
distances (defined as the minimum number of edges between two vertices) peaked
around a small characteristic value (i.e. {\em small world} \cite{Watts1998}
effect) and the degree (defined as the number of links per vertex) is
distributed according to a power law function.
Driven by the widespread presence of these common features, much work has been
done in order to understand the basic mechanisms underlying such universality
\cite{Barabasi2001a,Dorogovtsev2002,Caldarelli2002a,Newman2003b}. Instead, we
focus on a further characterization of networks using spectral methods that have
been used for detecting communities in complex networks
\cite{Hall1970a,Seary1995a,Kleinberg1999a,Capocci2004a}.  Community detection
techniques represent now a flourishing, inter-disciplinary field
\cite{Wu2004a,Radicchi2004a,Newman2003c,Capocci2004a,
Reichardt2004a,Fortunato2004a,Fontoura2004a,Donetti2004a}.  The aim of the
present paper is to show how spectral methods
\cite{Hall1970a,Seary1995a,Kleinberg1999a,Capocci2004a} cluster four classes of
content words: verbs, nouns, adverbs and adjectives in a Syntactic Dependency
Network \cite{Ferrer2003f} (SDN), which is a kind of linguistic network. The
topology of various instances of linguistic networks has recently been under
consideration by several studies. Examples include thesaurus networks based on
the Roget's thesaurus \cite{Steyvers2001,Motter2002,Newman2003b} and networks
based on Merrian-Webster's thesaurus \cite{Barabasi2001a}, WordNet
\cite{Steyvers2001,Sigman2001}, word association networks
\cite{Steyvers2001,Capocci2004a}, word co-occurrence networks
\cite{Ferrer2001a}, and SDNs \cite{Ferrer2003f}. The latter, as said above, will
be the target of the present article.
\begin{figure}
\begin{center}
\includegraphics[scale=0.8]{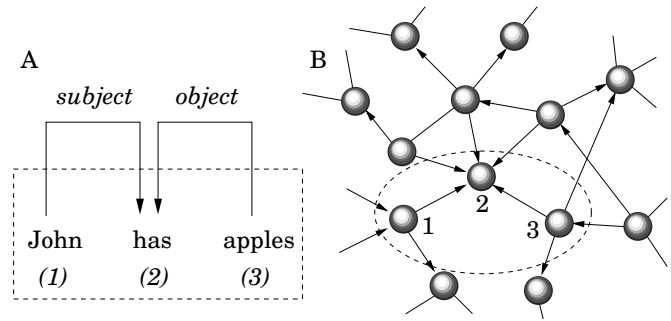}
\caption{\label{dependency_tree_figure}
A) The syntactic structure of a simple sentence. Words are the vertices 
and the syntactic dependencies are the edges of the graph. 
The proper noun 'John' and the verb 'has' are syntactically dependent 
in this sample sentence. 
'John' is {\em modifier} of the verb 'has', which is its {\em head}. 
Similarly, the action of 'has' is modified by its object 'apples'. 
Here we assume the graph oriented with edges pointing from a {\em modifier} to 
its {\em head}. 
B) Mapping the syntactic dependency structure of the sentence into a 
global syntactic dependency network.}
\end{center}
\end{figure}

SDNs are constructed by collecting the syntactic dependency links from a corpus,
i.e. a set of sentences.  Syntactic links are defined according to the
dependency grammar formalism \cite{Melcuk1988}, a special case of a broad family
of grammatical formalisms \cite{Hudson1984,Sleator1991}.  Dependency grammar
assumes that the syntactic structure of sentences consists of vertices (words)
and word pairwise connections (syntactic dependencies).  In this approximation,
a dependency link connects a pair of words representing an edge of the graph.
In the simple sentence ``John has apples'', ``John'' is linked to ``has'' and
``apples'' is linked to ``has''.  As explained in
Fig. \ref{dependency_tree_figure}, we can assign a direction to this
edges. ``John'' has the function to modify the meaning of the word ``has'' so
``John'' plays the role of {\em modifier} and the word ``has'' is referred as
{\em head}.  Most of edges are then directed and we assume that arrows go from
the modifier to its head (other conventions may make the opposite choice).
In some cases, such as in coordination, there is no clear direction
\cite{Melcuk2002}. But since such cases are rather uncommon, we will assume that
every link has a definite direction, by assigning an arbitrary direction to the
undirected cases.

We define a SDN as a set of $n$ words labeled with naturals from $1$ to
$n$. Syntactic dependencies are specified by the adjacency matrix
$A=\{a_{ij}\}$. $a_{ij}=1$ if the $i$-th word points to the $j$-th word, and
$a_{uj}=0$ otherwise. In fact, $a_{ij}=1$ are connected if the the $i$-th has
pointed to the $j$-th word at least once in a sentence from a corpus. The
syntactic dependency structure of a sentence can be seen as a subset of vertices
and links contained in a global network (Fig. \ref{dependency_tree_figure} B).
Syntactic dependencies between words in a sentence tend to be local and the
distance between syntactically linked words in a sentence decays exponentially
\cite{Ferrer2002e,Ferrer2004b}.  The organization of the rest of the papers can
be summarized as follows. Section \ref{community_detection_section} gives a
brief account of community detection techniques, with a special emphasis on
spectral methods and the way they will be applied here. Section
\ref{methods_section} describes the sources of the data and the statistical
measures that will be used to show how the spectral methods cluster words of the
same class. Section \ref{results_section} presents the results. A discussion of
the findings and some conclusions are reported in section
\ref{discussion_section}.

\section{The community detection technique}

\label{community_detection_section}

Various techniques for detecting community structure have been proposed recently
\cite{Wu2004a,Radicchi2004a,Newman2003c,Capocci2004a,Reichardt2004a,
Fortunato2004a,Fontoura2004a,Donetti2004a}.  Some of them
\cite{Radicchi2004a,Newman2003a} use the same arguments underlying the algorithm
introduced by Newman and Girvan \cite{Newman2003c} (NG--algorithm), and they
focus on the value of the edge betweenness. This quantity measures the fraction
of all shortest paths passing over a given link, or, according to an alternative
definition, the probability that a random walk on the network runs through that
link. By removing edges with large betweenness, one splits the whole network
(step by step) into connected components.  The process goes on until the whole
graph is decomposed in communities consisting of one single node each.  The
method is very efficient whenever some clues on the community structure is at
hand.  Otherwise, it does not give an indication of the resolution of the
clustering. Therefore it needs extra information in input (like the expected
number of clusters). Furthermore its outcome is independent on how sharp the
partitioning of the graph is.

To overcome such problems we adopted a different approach based on the spectral
analysis \cite{Capocci2004a}.  That approach conjugates the power of spectral
analysis with the caution needed to reveal an underlying structure when there is
no clear cut partitioning, as is in the case of the SDN considered.

Spectral methods are based on the analysis of simple functions of the adjacency
matrix $A$ \cite{Hall1970a,Seary1995a,Kleinberg1999a}.  In particular, the most
widely investigated function of $A$ are the Laplacian matrix $L=K-A$, the Normal
matrix $N = K^{-1}A$ and the composition $AA^T$, particularly useful when
dealing with oriented graphs. In the quantities defined above, we have assumed
that $K$ is the diagonal matrix with elements $k_{ii} = \sum_{j=1}^n a_{ij}$ and
$n$ is the number of nodes in the network. In most approaches concerning
undirected networks, $A$ is assumed to be symmetric, in contrast with the
present case, which explains our emphasis on the matrix $AA^T$. 

Just to familiarize the reader with the concept of the connection between
spectral analysis and communities, let us consider the simplest case among those
described above, the matrix $N$. Elements in a row can be interpreted as
probabilities that a walker moves from a given vertex to the other ones, since
those elements sum up to one. By such a probabilistic approach, it is evident
that the largest eigenvalue of the matrix $N$ is always equal to one. This
eigenvalue is associated to a trivial constant eigenvector, due to row
normalization.  In a network with an apparent cluster structure, $N$ has also a
certain number $m-1$ of eigenvalues close to one, where $m$ is the number of
well defined communities, the remaining eigenvalues lying a gap away from
one. We denote by $x_i$ the position of the $i$-th vertex after sorting vertices
by the value of its corresponding component in one eigenvector.  The
eigenvectors associated to these largest $m-1$ eigenvalues have a characteristic
structure too: the components corresponding to nodes within the same cluster are
assigned very similar $x_i$ values so that, as long as the partition is
sufficiently sharp, the profile of each eigenvector, sorted by components, is
step--like. The number of steps in the profile gives again the number $m$ of
communities \cite{Capocci2004a}.  A similar information is encoded in the
non-negative definite Laplacian matrix, where the eigenvalues close to zero are
associated to clusters.  While one can easily show that these spectral
properties are associated with the clustering pattern in the Normal matrix case,
this is less evident for the other cases. Nevertheless, it has been shown
\cite{Kleinberg1999a} that the spectral structure of the $AA^T$ helps in
detecting sets of highly mutually connected nodes in directed networks, thereby
indicating the presence of communities in the network such as the World Wide
Web. We will use the same empirical evidence here for the analysis of our word
dataset.

As explained in \cite{Capocci2004a}, solving the eigenvalue problem is
equivalent to minimization of a suitable function of the $x_i$s under a
suitable constraint.  The absolute minimum corresponds to the trivial
eigenvector, which is constant. The remaining stationary points correspond to
eigenvectors where components associated to well connected nodes assume similar
values.

For the present aim, it is enough to mention that (as for general real networks)
the typical eigenvector profiles in our SDN are not step-like, but rather
resemble a continuous curve.  Nevertheless, the method still applies. In fact,
we will see that components corresponding to nodes belonging to the same class
(or equivalently, that do not belong to a certain class) are still strongly
correlated and take, in each eigenvector, similar values among themselves. Thus,
a natural way to identify communities in an automatic manner is to measure
the correlation
\begin{equation}
r_{ij} = \frac{ \langle x_i x_j \rangle -
\langle x_i \rangle \langle x_j \rangle }
{[(\langle x_i^2 \rangle - \langle x_i \rangle^2)
(\langle x_j^2 \rangle - \langle x_j \rangle^2)]^\frac{1}{2}}\,,
\end{equation}
where the average $\langle \cdot \rangle$ is computed over the first few
nontrivial eigenvectors. The quantity $r_{ij}$ measures the community closeness
between the vertices $i$ and $j$. Though the performance may be improved by
averaging over a larger number of eigenvectors, with increased computational
effort, we found that indeed a small number of eigenvectors suffices to measure
the correlation between two vertices, which is positively correlated with the
chance that the two networks are strongly correlated \cite{Capocci2004a}.

The spectral method is used in a different manner here. The class each node
belongs to is known in advance; we check, then, whether the spectral method
clusters words of the same class, once they are sorted by their component in the
eigenvector. Since we know that the spectral method detects the community
structure, clusters are also likely to correspond to communities. 
With this aim,
we define $S=\{1,...,i,...,n\}$ as a sequence of vertices where $n$ is the
number of vertices of the syntactic dependency network. We assume that $S$ is
the outcome of taking one eigenvector and sorting the vertices by the value of 
the corresponding components in the eigenvector.  We define
$C=\{c_1,...,c_i,...,c_n\}$, a boolean vector indicating, for every vertex in
$S$, whether it can be of the class under consideration or it cannot. More
precisely, we have $c_i=1$ if the $i$-th vertex of $S$ can be of class $i$ and
$c_i=0$ otherwise.  The correlations we are interested in are not computed
between values in $X=\{x_1,...,x_i,...,x_n\}$ as in \cite{Capocci2004a} but
between values in $C$. 

\section{Methods}

\label{methods_section}

For the spectral analysis, we will consider two different matrices, i.e.  $A$
and $AA^T$. We use $AA^T$ because it is a way of obtaining equivalent
vertices. Here equivalence means playing the same role in the network,
i.e. pointing approximately to the same set of nodes. With this matrix, we are
following the same approach as in \cite{Kleinberg1999a}. $A$ is used for two
reasons: simplicity and as null hypothesis for $AA^T$.  The methods in
\cite{Hall1970a,Seary1995a} do not use directly $A$ but close variations such as
the Laplacian or the Normal matrix.  The main difference between the present
application of the spectral technique and the one performed in
\cite{Capocci2004a} are the following. First, we use simple adjacency matrices
instead of matrices with edge weights. Second, we do not normalize the
matrices. Normalizing the matrix is problematic here because most of verbs have
no outgoing link (recall the convention adopted here is that arcs go from
modifiers to heads). This would introduce special vertices in the graph without
outgoing links. This would affect the statistics of the whole system, and this
is why we did not normalize our adjacency matrix.  Besides, nodes without
outgoing links would mean trivial stationary solutions for a random walk on the
network. This means that a walker moving on the graph would systematically get
stuck in these ``sink'' vertices.  Another solution pursued in
\cite{Capocci2004a} is to remove vertices with no in--going or out--going links
(pruning). Nevertheless, here pruning would mean a drastic reduction of verbs
which may preclude the analysis of that class.
 
We introduce now various measures of the degree of clustering of words of a
given class in $C$. We consider a special case of $C$, $C_{scrambled}$, sharing
the same composition of $C$ but with scrambled components. The measures obtained
for $C_{scrambled}$ will be used as a null--hypothesis for those obtained for
$C$. Significant clustering cannot be claimed unless the clustering obtained
with $C$ and $C_{scrambled}$ differ clearly. Scrambled sequences have been used to test
the significance of long--range correlations in DNA sequences \cite{Li1992}. We
used the first non--trivial $98$ eigenvectors, so that we have $98$ $C$ vectors
for each class.

We define $k$ as a random variable measuring the length of a cluster of $1$s in
$C$. A cluster is a maximal sequence of consecutive elements equal to one in $C$. If
$C=\{0,0,0,1,1,1,0,0,0,0,1,0,1,1,1\}$, we have two clusters of length $3$
('$111$') and one cluster of length $1$ ('$1$').
We define $\left< k \right>$ as the mean value of $k$ in $C$. 

We define $P(k)$ as the probability that a cluster has size $k$. We will
estimate $P(k)$ from the proportion of clusters of length $k$ in $C$. $k$ is
expected to follow a geometric distribution in $C_{scrambled}$. There, we have
\begin{equation}
P(k)=\left<c_i\right>^k (1-\left<c_i\right>),
\label{geometric_distribution_equation}
\end{equation}
where $\left<c_i \right>$ is the mean value of the components of
$C$. In other words, $\left<c_i \right>$ is the proportion of ones of
$C$.
From equation \ref{geometric_distribution_equation}, it follows that 
\begin{equation}
\left<k\right>=1/(1-\left<c_i\right>)
\end{equation}
 and the standard deviation is 
\begin{equation}
\sigma(k)=\left<c_i \right>^{1/2}/(1-\left<c_i \right>).
\end{equation}
Since the estimated $P(k)$ may contain a considerable 
amount of noise, we will use, $P_\geq(k)$, the cumulative $P(k)$, defined as 
\begin{equation}
P_\geq(k)=\sum_{K\geq k}^\infty P(K).
\end{equation}

We are also interested in measures of the correlations between $c_i$ and
$c_{i+d}$, with $d>0$. 
If the community detection clusters words of the same community
consecutively in $C$ \cite{Capocci2004a}, correlations above the expected value
in the scrambled sequence are expected.  Here we will use two measures of
correlation: $\Gamma(d)$ and $I(d)$, that are, respectively, the Pearson
correlation coefficient and the information transfer, between vertices at
distance $d$ in $C$ \cite{Li1992}.  $\Gamma(d)$ and $I(d)$ are complementary
measures. $\Gamma(d)$ measures positive and negative correlations, whereas
$I(d)$ cannot distinguish positive from negative correlations. $I(d)$ captures
non--linear correlations that $\Gamma(d)$ does not detect \cite{Li1990}. $I(d)$
is apparently more sensitive to finite sampling than $\Gamma(d)$. Here,
$\Gamma(d)$ is defined as
\begin{equation}
\Gamma(d)=\frac{COV(c_i,c_{i+d})}{\sigma(c_i) \sigma(c_{i+d})},
\label{raw_correlation_equation}
\end{equation}
where $1 \leq d<n$, 
$COV(c_i,c_{i+d})$ is the covariance between $c_i$ and $c_{i+d}$,
$\sigma(c_{i+d})$ is the standard deviation of $c_{i+d}$ and
$\sigma(c_i)=\sigma(c_{i+d})$ with $d=0$.
We have 
\begin{equation}
\left<c_{i+d}\right>=\frac{1}{n-d}\sum_{i=1}^{n-d}c_i.
\end{equation}
The covariance is defined as 
\begin{equation}
COV(c_i,c_{i+d})=\sum_{i=1}^{n-d}\frac{(c_i-\left< c_i
  \right>)(c_{i+d}-\left< c_{i+d} \right>)}{n-d}.
\label{covariance_equation}
\end{equation}
 The standard deviation is defined as 
\begin{equation}
\sigma(c_{i+d})=(\left< c^2_{i+d} \right> - \left<c_{i+d} \right>^2)^{1/2}
\end{equation}
as usual.
Given that $\left< c_i^2 \right>= \left< c_i \right>$ because $C$ is a binary sequence, it
follows
\begin{equation}
\sigma(c_{i+d})=(\left< c_{i+d} \right>(1 - \left<c_{i+d}
\right>))^{1/2}.
\label{standard_deviation_equation}
\end{equation}
Replacing equations \ref{covariance_equation} and
\ref{standard_deviation_equation} in equation \ref{raw_correlation_equation}, we
obtain
\begin{equation}
\Gamma(d)=\sum_{i=1}^{n-d}\frac{(c_i-\left< c_i \right>)(c_{i+d}-\left< c_{i+d}
\right>)}{(n-d)((\left< c_i \right>(1 - \left<c_i \right>)(\left< c_{i+d}
\right>(1 - \left<c_{i+d} \right>))^{1/2}}.
\end{equation}

Here, $I(d)$ is defined as
\begin{equation}
I(d)=\sum_{x,y \in \{0,1\}} p(c_i=x,c_{i+d}=y)\log \frac{p(c_i=x,c_{i+d}=y)}{p(c_i=x)p(c_{i+d}=y)},
\end{equation}
where
\begin{equation}
p(c_i=x,c_{i+d}=y)=\frac{1}{n-d}\sum_{\begin{array}{c} i=1 \\
    c_i=x~and~c_j=y \end{array}}^{n-d} 1, 
\end{equation}
\begin{equation}
p(c_i=x)=\sum_{y \in \{0,1\}} p(c_i=x,c_{i+d}=y)
\end{equation}
and
\begin{equation}
p(c_{i+d}=y)=\sum_{x \in \{0,1\}} p(c_i=x,c_{i+d}=y).
\end{equation}

\begin{table}
\begin{center}
\begin{tabular}{lrr}
Type       & Number & Proportion \\ \hline
Verbs      & $985$  & $0.17$ \\
Nouns      & $3093$ & $0.55$ \\ 
Adverbs    & $171$  & $0.03$ \\
Adjectives & $1129$ & $0.20$ \\ 
Other      & $337$  & $0.06$ \\              
\end{tabular}
\caption{\label{category_table}
  Counts of the frequency of very word class. The total
  amount of different words (i.e. vertices in the syntactic dependency
  network) is $5563$. A word counts for a certain class if it has
  appeared at least once for that class in the Romanian corpus
  studied in \cite{Ferrer2003f}. Since a word can be of different
  classes, the sum of the column 'Number' does not necessarily
  equals the number of different words. }
\end{center}
\end{table}

Now we study the Romanian syntactic dependency network described
in \cite{Ferrer2003f} with $n=5563$ vertices. We choose that network from the
set of three networks studied in \cite{Ferrer2003f} because it is the most complete. The other networks systematically lack several syntactic
dependencies.  The network is a small-world one with significantly high
clustering, has a power--law distribution of vertex degrees and exhibits
disassortative mixing.  Those properties are shared with other non-linguistic
biological networks \cite{Ferrer2003f}. As in \cite{Ferrer2003f}, we worked on
the largest connected component.  A given word belongs to a given class if that
word is labeled under that class at least once in the corpus that originated the
syntactic dependency network in \cite{Ferrer2003f}. Table \ref{category_table}
shows the number of words in each class according the previous
definition.  Only $6\%$ of words cannot fall in any of the previous classes.
Participles and infinitives where excluded from the class verb, as it was done
in the original corpus. Those words represent a very small fraction of the
vertices.

\section{Results}

\label{results_section}

\begin{figure*}
\begin{center}
\includegraphics[scale=0.6]{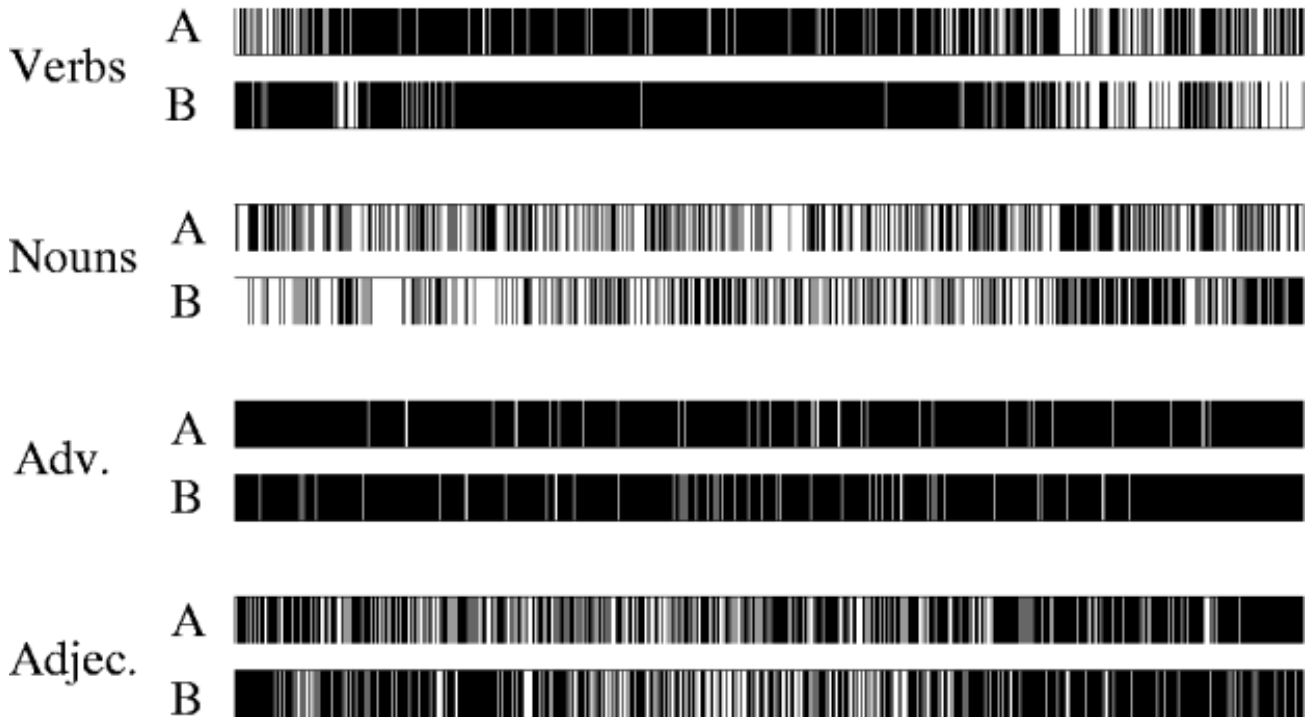}
\caption{\label{communities_1D_figure}
  Visual representation of the some binary vectors $C=\{c_i\}$
  delivered by the spectral methods for different word
  classes: verbs, nouns, adverbs and adjectives. 
  $c_i=1$ if the $i$-th vertex (or word) belongs to the
  class under consideration and $c_i=0$, otherwise. $C$ has $5563$
  components, which is the number of vertices of the SDN.
  $C$ is the
  outcome of ordering vertices by the the corresponding 
  values of the eigenvectors delivered by the spectral methods and replacing 
the vertex by 1 if it is of the class under
  consideration or by 0 otherwise. 
  White and black indicate, respectively, $c_i=1$ and
  $c_i=0$. Two types of $C$ are shown for each word class: (A) an ordering
  using $A$ and (B) ordering using $AA^T$.}
\end{center}
\end{figure*}

Fig. \ref{communities_1D_figure} shows some examples of $C$ for nouns,
adverbs and adjectives. Large clusters can be visually identified.
Table \ref{cluster_size_statistics_figure} gives mean $\left< k \right>$ and
the maximum value of the cluster size over the sample set of
eigenvectors.

\begin{table}
\begin{center}
\begin{tabular}{llccc}
Class      & Cluster size & $A$ & $AA^T$    & Scrambled  \\ \hline \hline
Verbs      & $\left< k \right>$ & $2.01$    & $2.08 \pm 0.32$    & $1.21 \pm 0.51$  \\ 
           & $max$  & $43$      & $24.96 \pm 20.87$ & $4.72 \pm 0.84$  \\ \hline
Nouns      & $\left< k \right>$ & $2.66$    & $2.68 \pm 0.13$   & $2.25 \pm 1.67$  \\ 
           & $max$  & $32$      & $34.35 \pm 10$    & $13.73 \pm 2.04$ \\ \hline  
Adverbs    & $\left< k \right>$ & $1.06$    & $1.08$            & $1.03 \pm 0.18$ \\ 
           & $max$  & $2$       & $3$               & $2.16 \pm 0.39$ \\ \hline
Adjectives & $\left< k \right>$ & $1.39$    & $1.53$            &
$1.25 \pm 0.56$          \\ 
           & $max$  & $10$      & $8.14 \pm 0.83$   & $5.11 \pm 0.80$  
\end{tabular}
\caption{\label{cluster_size_statistics_figure} Cluster sizes for the
  four classes of words, the two types of matrices considered ($A$ and
  $AA^T$) and scrambled vertex orderings. Standard errors smaller than $10^{-2}$ are not shown. }
\end{center}
\end{table}

\begin{figure*}
\begin{center}
\includegraphics[scale=0.6]{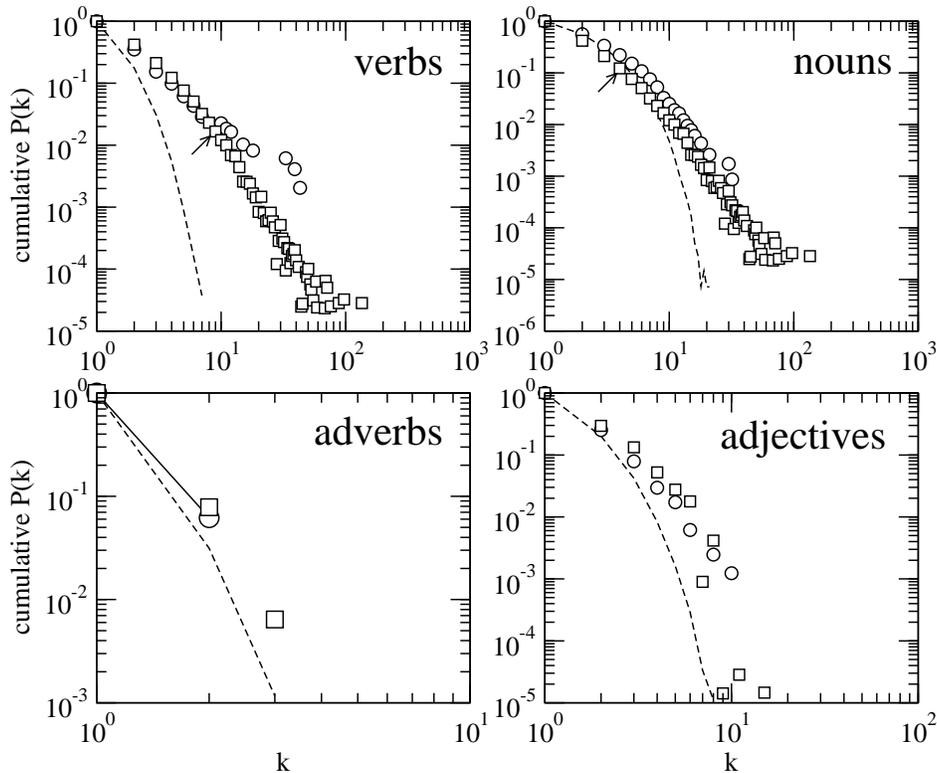}
\caption{\label{cluster_size_probability_figure}
  Cumulative $P(k)$ where $P(k)$ is the proportion of clusters of
  length $k$. Results using $A$ (circles), $AA^T$
  (squares) and the null hypothesis (dashed line) are shown. 
  A clear power trend is found for the series of $AA^T$ for verbs and
  nouns from the each arrow to the right. }
\end{center}
\end{figure*}

Fig. \ref{cluster_size_probability_figure} shows the mean $P(k)$ over
the sample set of eigenvectors for the two kinds
of matrices and the null hypothesis.
For verbs and nouns, $P(k)$ obeys
\begin{equation}
P(k) \sim k^{-\gamma k} 
\label{power_distribution_equation}
\end{equation}
for sufficiently large $k$.
The agreement with equation \ref{power_distribution_equation} 
is lower adopting $A$ rather than $AA^T$. 
Verbs and nouns cannot be given account of by the geometric distribution
predicted by the null hypothesis, supporting thus the significance of the
results. No conclusion can be made on adverbs, since they are not represented
enough and it seems unlikely that adjectives follow equation
\ref{power_distribution_equation}.

\begin{figure*}
\begin{center}
\includegraphics[scale=0.6]{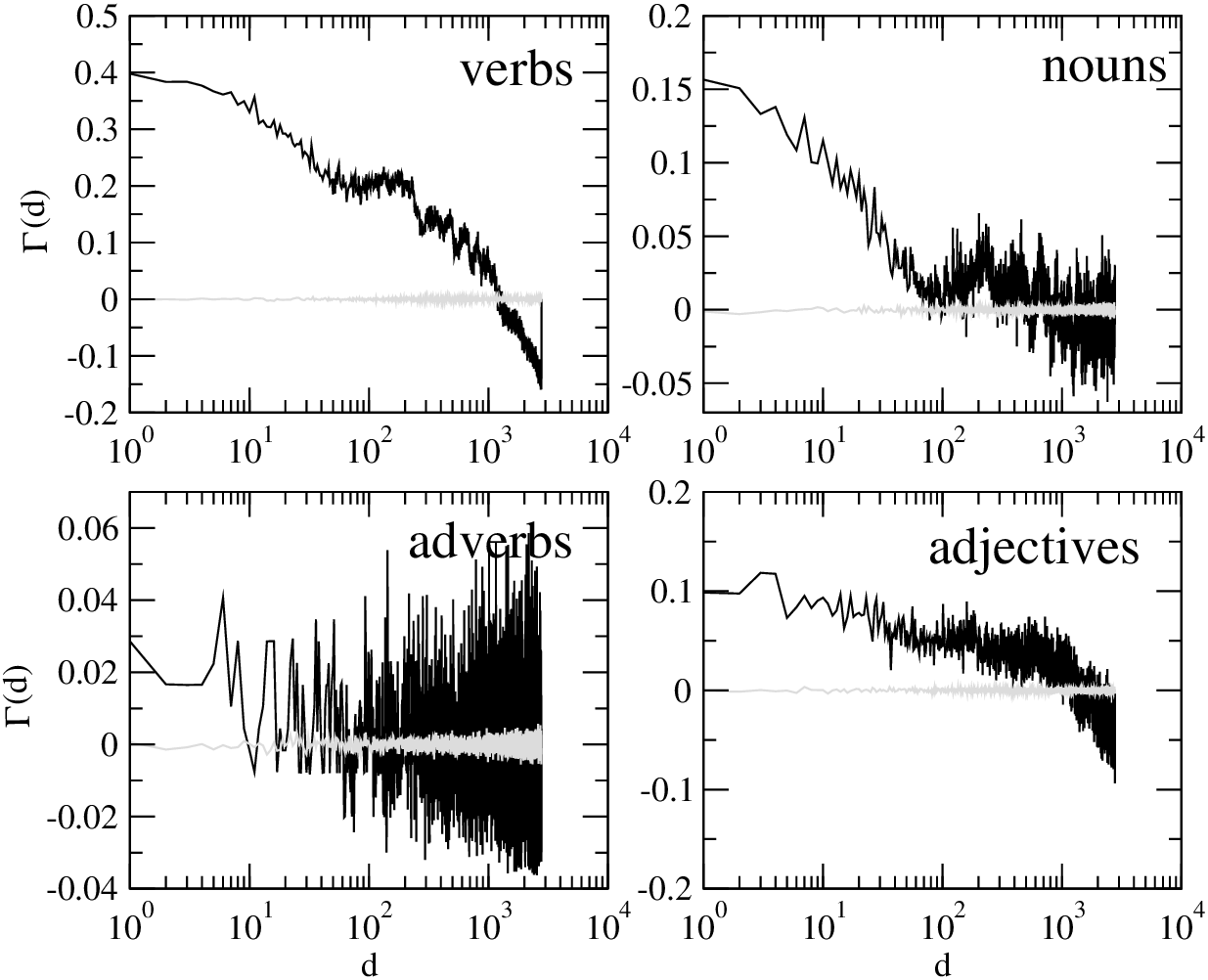}
\caption{\label{correlation_A_figure} $\Gamma(d)$, the correlation
  coefficient between words of a particular class as a function of $d$, the
  distance in a vertex ordering provided by the spectral methods. 
  Two series are shown for each word class: $\Gamma(d)$
  using $A$ (black) and the $\Gamma(d)$ for the scrambled ordering (gray).}
\end{center}
\end{figure*}

\begin{figure*}
\begin{center}
\includegraphics[scale=0.6]{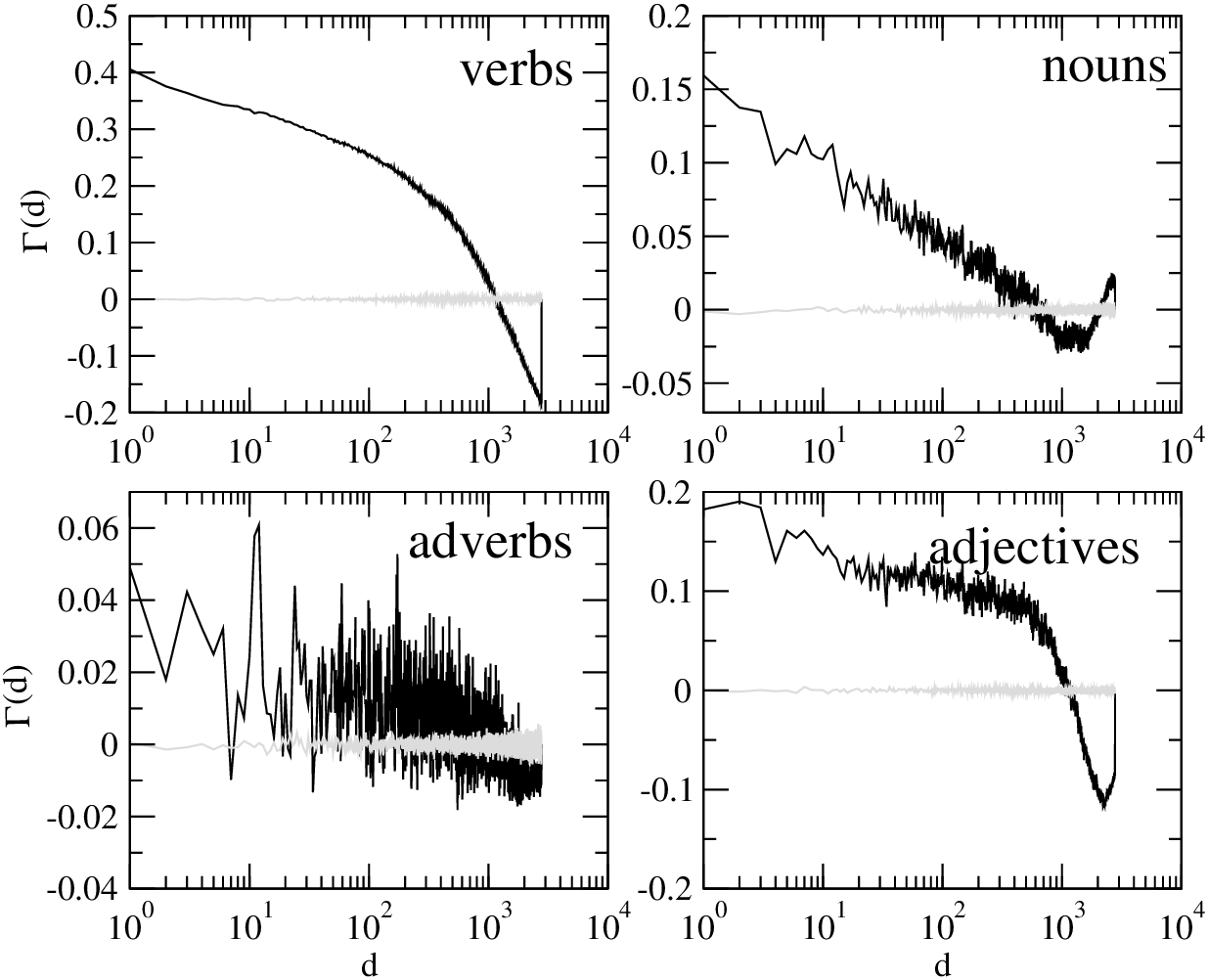}
\caption{\label{correlation_AAT_figure} $\Gamma(d)$, the correlation
  coefficient between words of a particular class as a function of $d$, the
  distance in a vertex ordering provided by the spectral methods. 
Two series are shown for each word class: $\Gamma(d)$
  using $AA^T$ (black) and the $\Gamma(d)$ for the scrambled ordering (gray).}
\end{center}
\end{figure*}

\begin{figure*}
\begin{center}
\includegraphics[scale=0.6]{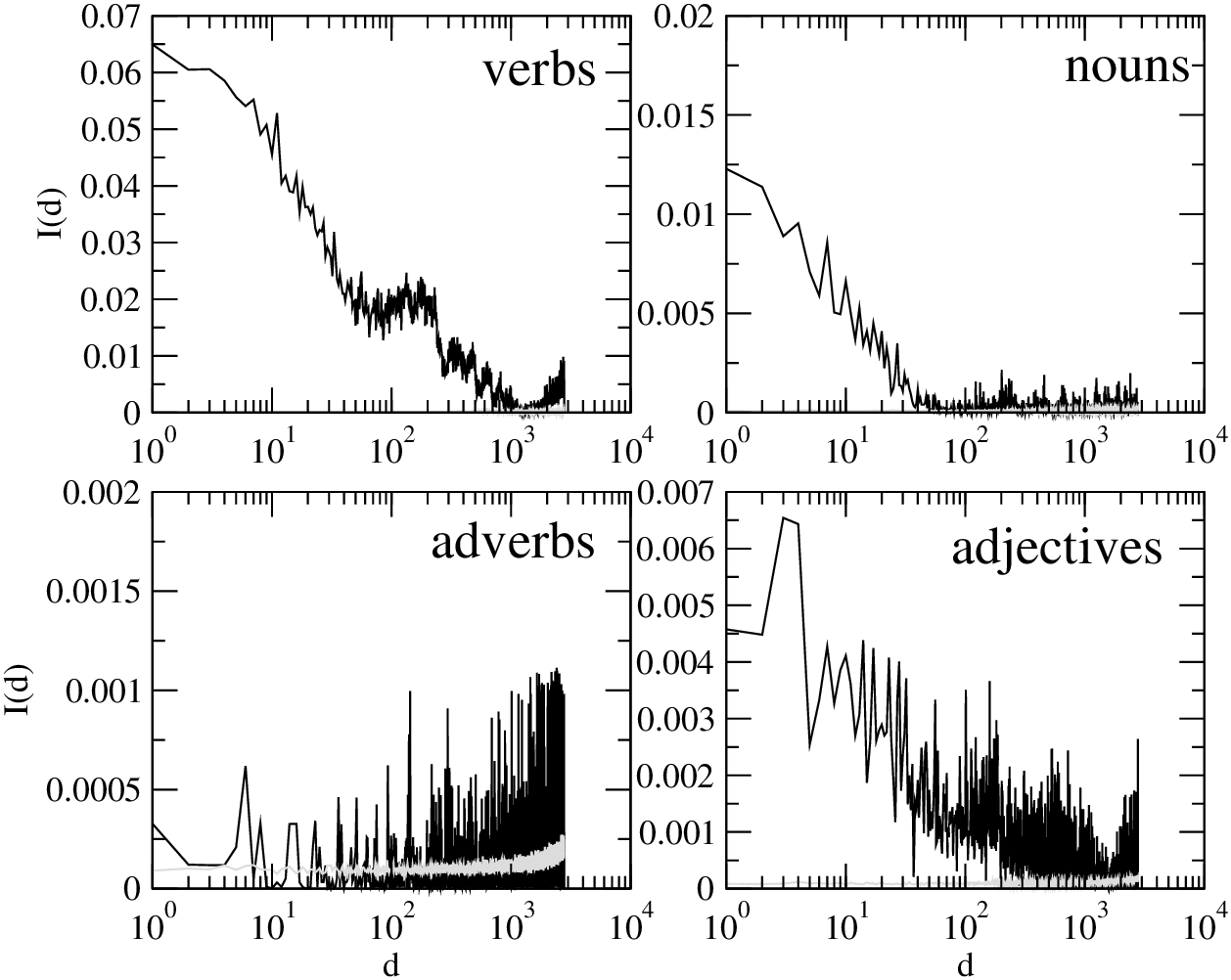}
\caption{\label{information_A_figure} $I(d)$, the information
  between words of a particular class as a function of $d$, the
  distance in a vertex ordering provided by the spectral methods. 
Two series are shown for each word class: $I(d)$
  using $A$ (black) and $I(d)$ for the scrambled ordering (gray).}
\end{center}
\end{figure*}

\begin{figure*}
\begin{center}
\includegraphics[scale=0.6]{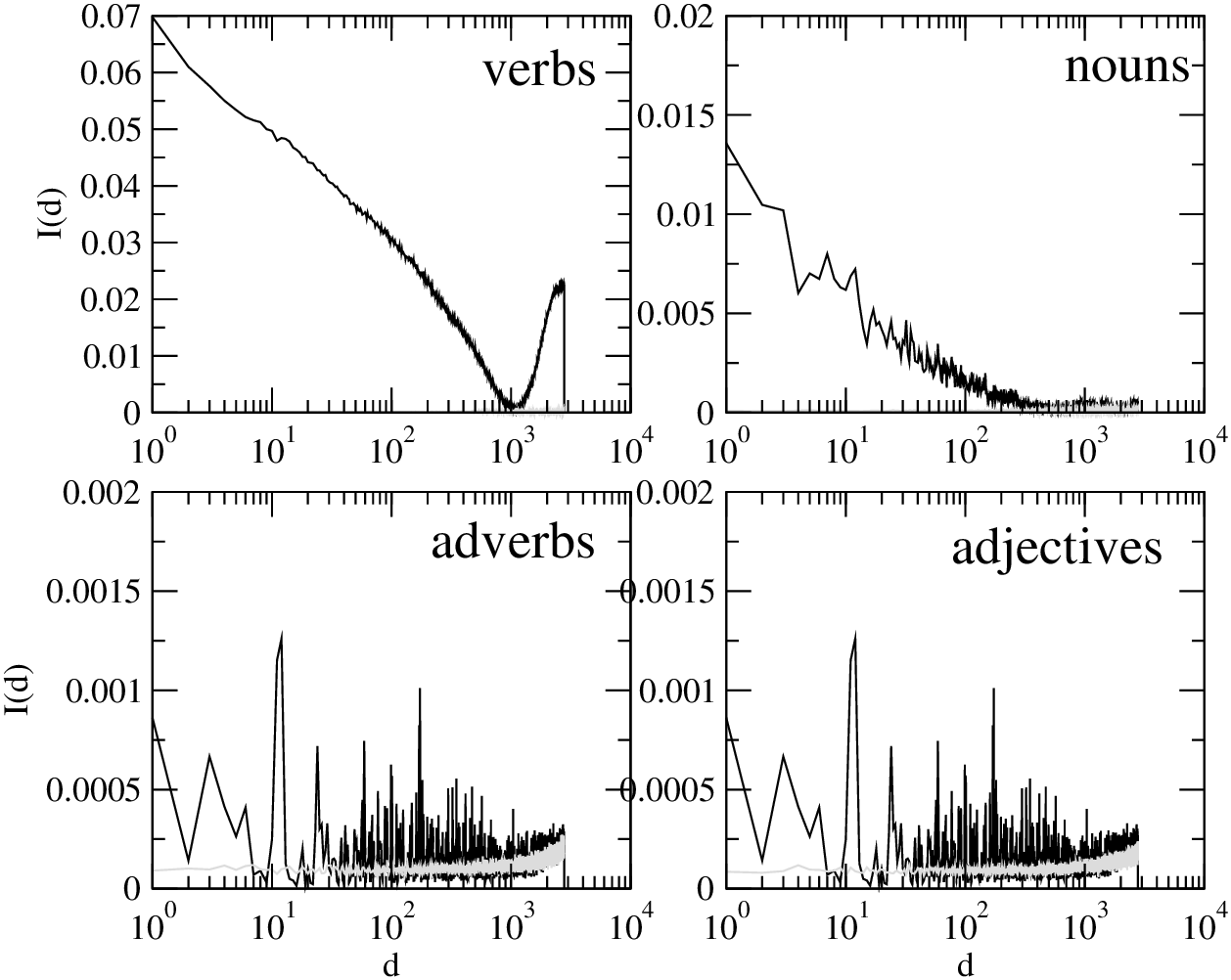}
\caption{\label{information_AAT_figure} $I(d)$, the information
  between words of a particular class as a function of $d$, the
  distance in a vertex ordering provided by the spectral methods. 
Two series are shown for each word class: $I(d)$
  using $AA^T$ (black) and $I(d)$ for the scrambled ordering (gray).}
\end{center}
\end{figure*}

Figure \ref{correlation_A_figure} and \ref{correlation_AAT_figure} show
$\Gamma(d)$ for $A$ and $AA^T$, respectively. 
Fig. \ref{information_A_figure} and
\ref{information_AAT_figure} show $I(d)$ for $A$ and $AA^T$,
respectively.  
In order to determine
the length of the correlations for a certain correlation measure in a
conservative way, we define two series: $B_L(d)$ and $B_U(d)$. $B_L(d)$ is the
mean value of a real correlation for distance $d$ over the sample of
eigenvectors minus the corresponding standard deviation. $B_L(d)$ is an
approximate lower bound for the real value of the correlation.  $B_U(d)$ is the
mean value of the null hypothesis series over the sample of eigenvectors times
the corresponding standard deviation. $B_U(d)$ is an approximate upper bound for
the null hypothesis.  Since real correlations tend to decrease with length
(Figs. \ref{correlation_A_figure}, \ref{correlation_AAT_figure} and
\ref{information_A_figure} and \ref{information_AAT_figure}), an approximate
conservative measure of the length of statistically significant correlations is
$d^*$, the smallest value of $d$ at which $B_L(d)$ and $B_U(d)$
cross. Long-distance correlations where found (Table
\ref{length_of_statistically_significant_correlations_table}), except for
adverbs, due to the small amount of vertices that can be adverbs (recall Table
\ref{category_table}).

\begin{table}
\begin{center}
\begin{tabular}{llrr}
Class      & Correlation &     & $d^*~~~$  \\ \hline
           &             & $A$ & $AA^T$ \\ \hline \hline
Verbs      & $\Gamma(d)$ & $1164$  & $915$ \\
           & $I(d)$ & $872$  & $683$ \\ \hline
Nouns      & $\Gamma(d)$ & $50$  & $63$ \\
           & $I(d)$ & $37$  & $13$ \\ \hline
Adverbs    & $\Gamma(d)$ & $6$  & $6$ \\
           & $I(d)$ & $1$  & $1$ \\ \hline
Adjectives & $\Gamma(d)$ & $148$  & $881$ \\
           & $I(d)$ & $36$  & $648$ \\ \hline
\end{tabular}
\caption{\label{length_of_statistically_significant_correlations_table} 
   $d^*$, the approximate length of statistically significant
   correlations using two different correlation measures measures: the Pearson
   correlation coefficient $\Gamma(d)$ at distance $d$ and the
   information transfer at distance $I(d)$. Four classes of words are considered.}
\end{center}
\end{table}

\section{Discussion}

\label{discussion_section}

In the previous sections we have obtained three main results.  First, the
spectral methods clusters significantly words of the same class. Second, those
spectral methods sort vertices in a way such that long-range correlations appear
in the binary vector of membership to a certain class. Third, $AA^T$ clusters
word classes better than $A$. The chances of getting a large cluster are higher
using $AA^T$ (Fig. \ref{cluster_size_probability_figure}).  The results
presented here confirm the power of the spectral methods introduced in
\cite{Capocci2004a} for detecting community structure in linguistic
networks. 
Using a priori information about class membership of every vertex, 
we have discovered
that the spectral method clusters words consistently with the word types that
linguists have been distinguishing from a long time ago.

Discovering word classes minimizing the {\em a priori} amount of linguistic
knowledge is a challenge in linguistics. Recently, Montemurro \& Zanette
\cite{Montemurro2002b} have clustered words using only information about the
degree of heterogeneity with which words are distributed throughout a text. In
particular, they have found that nouns and adjectives tend to be more
heterogeneously distributed than verbs and adverbs. In a syntactic network,
we have found that verbs and
nouns and significantly more heterogeneously distributed according to
the ordering provided by spectral methods than adverbs and
adjectives. The present work suggests that word classes could be
eventually discovered using only the structure of syntactic interactions.

\begin{acknowledgments}

We thank Ricard Sol\'e for the drawing in Fig. \ref{dependency_tree_figure}. 
This work was supported by the FET Open Project COSIN
(IST-2001-33555) and Integrated Project DELIS.

\end{acknowledgments}


\end{document}